\documentclass[aps,prc,twocolumn,groupedaddress,nofootinbib]{revtex4-2}
\RequirePackage{plautopatch}
\usepackage{bm}
\usepackage{graphicx}
\usepackage{ascmac}
\usepackage{amsmath,amssymb}
\usepackage{cancel}
\usepackage{braket}
\usepackage{url}
\usepackage[setpagesize=false,hidelinks,pdfusetitle]{hyperref}
\usepackage{mathtools}
\DeclarePairedDelimiter{\abs}{\lvert}{\rvert}
\allowdisplaybreaks[1]

\begin{document}


\title{Hadron-hadron potentials coupled to quark degrees of freedom for exotic hadrons}


\author{Ibuki Terashima}
\email[]{terashima-ibuki@ed.tmu.ac.jp}
\affiliation{Department of Physics, Tokyo Metropolitan University, Hachioji 192-0397, Japan}
\author{Tetsuo Hyodo}
\email[]{hyodo@tmu.ac.jp}
\affiliation{Department of Physics, Tokyo Metropolitan University, Hachioji 192-0397, Japan}


\date{\today}

\begin{abstract}
    We study the properties of the hadron-hadron potentials and quark-antiquark potentials from the viewpoint of the channel coupling. We demonstrate that, for finite quark masses, the coupling to the two-hadron continuum induces the imaginary part of the quark-antiquark potential, in contrast to the string-breaking phenomena in the static limit. It is also shown that the elimination of the different degrees of freedom induces the nonlocality and energy dependence of the effective potentials. For the obtained nonlocal potentials, we apply two methods of the local approximation proposed previously, the formal derivative expansion and the derivative expansion in the HAL QCD method, by carefully examining the energy dependence of the potential. As an application, we construct a coupled-channel model of $c\bar{c}$ and $D^{0}\bar{D}^{*0}$ to describe $X(3872)$, and discuss the property of the effective $D^{0}\bar{D}^{*0}$ potentials. We confirm that the local approximation by the HAL QCD method works better than the formal derivative expansion also for the energy-dependent potential. At the same time, we show that, in the HAL QCD method, the resulting phase shift is sensitive to the choice of the wave function to construct the local potential when the system has a shallow bound state such as $X(3872)$. 
    \end{abstract}


\maketitle

\section{Introduction}\label{sc_intro}
Focusing on the potentials helps us to reveal an essential mechanism behind the interactions between hadrons. For example, the basic properties of the nuclear force were known in the early 1930s, but the physical mechanism of the nuclear force was not clear until Yukawa suggested the $\pi$ exchange potential to give a physical picture for the nuclear force~\cite{Yukawa:1935xg}.
As a result of further developments of researches, the realistic nuclear forces that reproduce experimental observables with high accuracy were established~\cite{Wiringa:1994wb,Machleidt:2000ge}, and the nuclear force potentials by chiral effective field theory were also constructed with similar precision~\cite{Epelbaum:2008ga,Machleidt:2011zz,Hammer:2019poc}.
Nowadays, the strong interaction is described by quantum chromodynamics (QCD), and it is known that hadrons are composed of quarks. The studies of hadron interactions based on QCD are actively carried out by numerical calculations using lattice QCD~\cite{Ishii:2006ec,Aoki:2009ji,Aoki:2012tk,Aoki:2020bew}.

One way to study the color confinement, a nontrivial property of QCD, is to examine the potentials between the infinitely heavy static quarks ($Q\bar{Q}$ potentials).
For instance, by using the analytic methods of the strong coupling expansion~\cite{Wilson:1974sk}, it has been found that the linear behavior of the quark potential characterizes the color confinement. The authors of Ref.~\cite{Eichten:1974af} proposed the Cornell potential (Coulombic at short distances and linear at long distances) for the heavy quark potential.
In fact, the charmonium spectrum has been successfully reproduced by the Cornell potential~\cite{Godfrey:1985xj}.
Currently, the Cornell potential is also obtained by the numerical calculations by lattice QCD~\cite{Bali:2000gf}.

In this way, researches of quark potentials and hadron potentials have been developed independently in a framework with its own degrees of freedom.
However, since there is no restriction in QCD to prohibit the mixing of states with the same quantum numbers, the study of potentials with mixing of quark and hadron degrees of freedom should be important for understanding the strong interaction.
In fact, the internal structure of the exotic hadron $X(3872)$ discovered in 2003~\cite{Belle:2003nnu} is considered to be a mixture of the $c\bar{c}$ core state and the $D\bar{D}^*$  molecule state~\cite{Takizawa:2012hy,Hosaka:2016pey}. Even now, theoretical studies are intensively performed toward the complete elucidation of the internal structure of $X(3872)$~\cite{Guo:2017jvc,Brambilla:2019esw}.

As a previous study of potentials with such channel coupling, there is a lattice QCD study that succeeded in describing the string breaking of the $Q\bar{Q}$ potential in the static limit with the infinite quark mass~\cite{Bali:2005fu}. By calculating the $Q\bar{Q}$ potential with dynamical light quarks, it was shown that when the potential reaches the meson-meson threshold, a pair creation of light quarks occurs between the static quarks so that the potential is flattened at large distance.
Meanwhile, for more realistic quark potentials with a finite mass ($q\bar{q}$ potentials), lattice QCD calculations have been performed. It is shown that the $q\bar{q}$ potentials are of the Coulomb plus linear form~\cite{Ikeda:2011bs,Kawanai:2011xb,Nochi:2016wqg}.
Nevertheless it is not clear how the $q\bar{q}$ potential is affected by the channel coupling induced by the quark pair creations.

In this study, we construct a framework where the quark degrees of freedom couples to the hadron degrees of freedom, and consider the contribution of channel coupling in the effective potentials away from the static limit. As an application of the hadron-hadron effective potential that includes the contribution of the quark degrees of freedom, we discuss the properties of the effective $D\bar{D}^*$ interaction for the exotic hadron $X(3872)$ by using numerical calculations.

In this paper, we first formulate the Hamiltonian that describes the channel coupling problem in Sec.~\ref*{sub_sc_channel}, and derive the effective potential that contains the contribution from different degree of freedom. Next, in Sec.~\ref*{sub_sc_local}, we introduce the approximate transformation methods from the obtained effective nonlocal potential to a local one: the formal derivative expansion and the derivative expansion by the HAL QCD method~\cite{Aoki:2021ahj}. Furthermore, in Sec.~\ref*{sub_sc_yukawa}, we analytically derive the effective hadron-hadron potential with the Yukawa-type transition form factor with energy-dependent strength. In Sec.~\ref*{sc_Application}, we apply the approximated local effective potentials to $X(3872)$ in the $D\bar{D}^*$ scattering. First, in Sec.~\ref*{sub_sc_form_X}, we construct a model of $X(3872)$, and then discuss the results obtained by numerical calculations in Sec.~\ref*{sub_sc_nmrcl_result}. Finally, we summarize this study in Sec.~\ref*{sc_summary}. Preliminarily, results of the content of Sec.~\ref*{sub_sc_channel} are reported in the conference proceedings~\cite{Terashima:2022pba}.

\section{Formulation}\label{sc_formu}
\subsection{Effective potential with channel coupling}\label{sub_sc_channel}
In this section, we discuss the effective potentials between quarks and hadrons using the channel coupling. Deriving analytical expressions, we study the qualitative properties of the obtained effective potentials.
The channel-coupling problem between quark and hadron degrees of freedom is formulated in non-relativistic quantum mechanics. In this study, we follow the Feshbach's method~\cite{Feshbach:1958nx,Feshbach:1962ut} to describe the  coupled channel system. The Hamiltonian $H$ that couples the quark-antiquark channel with a confinement potential $V^q$ and the two-hadron channel with a scattering potential $V^h$ by a transition potential $V^t$ is given by
\begin{align}\label{eq_2ch_hamiltonian}
    H=
    \begin{pmatrix}
        T^{q} & 0             \\
        0     & T^{h}+ \Delta
    \end{pmatrix}
    +
    \begin{pmatrix}
        V^{q} & V^t   \\
        V^t   & V^{h}
    \end{pmatrix},
\end{align}
where $T^{q}$ and $T^{h}$ are the kinetic energies of the quark and hadron channels, respectively, and $\Delta$ is the threshold energy of the hadron channel. Then, the Schr\"{o}dinger equation is given by
\begin{align}\label{eq_2ch_schrodinger}
    H \ket{\Psi} =  E \ket{\Psi},
\end{align}
where the wave function $\ket{\Psi}$ is given by
\begin{align}
    \ket{\Psi}=
    \begin{pmatrix}
        \ket{q} \\
        \ket{h}
    \end{pmatrix},
\end{align}
and $\ket{q}$ and $\ket{h}$ are the wave functions of the quark and hadron channels, respectively.

By eliminating the quark channel, we obtain an effective Hamiltonian 
\begin{align}
    H^{h}_\mathrm{ eff} (E) & =  T^h + \Delta +V^{h}_\mathrm{{eff}}(E),
\end{align}
where we define the effective potential between hadrons as
\begin{align}\label{eq_Veff_h}
    V^{h}_\mathrm{{eff}}(E)=V^{h}+V^tG^{q}(E)V^t ,
\end{align}
and the Green's function for the quark channel as
\begin{align}
    G^{q}(E) = \left[E-(T^{q} + V^{q})\right]^{-1}.
\end{align}
$ H^{h}_\mathrm{ eff} (E)$ satisfies the Schr\"{o}dinger equation for the hadron channel only
\begin{align}\label{eq_sch_h}
    H^{h}_\mathrm{ eff} (E) \ket{h}= E \ket{h},
\end{align}
The second term in Eq.~\eqref{eq_Veff_h} is the potential due to the coupling with the quark channel.

Similarly, by eliminating the hadron channel, we obtain an effective Hamiltonian for the quark channel
\begin{align}
    H^{q}_\mathrm{ eff} (E)
     & = T^q +V^{q}_\mathrm{{eff}}(E),
\end{align}
and the effective potential between quarks
\begin{align}
    V^{q}_\mathrm{{eff}}(E)=V^{q}+V^tG^{h}(E)V^t,
\end{align}
where we define the Green's function for the hadron channel as 
\begin{align}
    G^{h}(E)=\left[E-(T^{h}+\Delta+V^{h})\right]^{-1}.
\end{align}
$H^{q}_\mathrm{ eff} (E)$ satisfies the Schr\"{o}dinger equation for the quark channel only 
\begin{align}\label{eq_sch_q}
    H^{q}_\mathrm{ eff} (E) \ket{q}= E \ket{ q}.
\end{align}

Since we have not made any approximations in the formulation of the effective potentials, we can obtain wave functions $\ket{q}$ and $\ket{h}$ that are equivalent to the original channel-coupled Schr\"{o}dinger equation~\eqref{eq_2ch_schrodinger} by solving Eqs.~\eqref{eq_sch_h} and~\eqref{eq_sch_q}.
However, it should be noted that in order to solve Eq.~\eqref{eq_sch_h}, we need to have obtained the explicit form of the Green's function for the quark channel $G^{q}(E)$. In other words, we cannot obtain the effective potential between hadrons without solving the quark channel.

To obtain a specific form of Green's functions, we first consider a system with only quark channels. The Schr\"{o}dinger equation for the quark channel in the absence of the channel coupling is 
\begin{align}
    (T^q + V^q)\ket{\phi} =E\ket{\phi},
\end{align}
where $\ket{\phi}$ is the wave function.
Since $V^q$ is a confinement potential, there are no scattering solutions, and the eigenstates are given only by the states $\ket{\phi_n}$ with discrete eigenvalues $E_n$ with $n=0,1,2,\cdots$. The state vector $\ket{\phi_n}$ is defined such that the normalization and orthogonality are given by
\begin{align}\label{eq_phi_kikaku}
    \braket{\phi_n \mid \phi_m}=\delta_{mn}.
\end{align}
Furthermore, since there are no eigenstates other than $\ket{\phi_n}$, from the completeness of the basis
\begin{align}\label{eq_phi_kanzen}
    \sum_n \ket{\phi_n}\bra{\phi_n} = 1^q,
\end{align}
the unit matrix $1^q$ of the quark channel is expressed by $\ket{\phi_n}$.

To investigate the properties of the effective hadron potential coupled with the quark channels in the coordinate representation, we sandwich the effective hadron potential~\eqref{eq_Veff_h} with the eigenvector $\ket{\bm{r}_{h}}$ of the position operator of the hadron channel to obtain
\begin{align}
    \braket{\bm{r'}_{h} \mid V^{h}_\mathrm{ eff}(E)\mid\bm{r}_{h}}
    =                                                                                  \braket{\bm{r'}_{h} \mid V^{h}\mid\bm{r}_{h}} +\sum_n \frac{F^*_n(\bm{r'}) F_n(\bm{r})}{E-E_n}. \label{eq_VDD}
\end{align}
Here, we use the complete set of the basis Eq.~\eqref{eq_phi_kanzen}, and we define the form factor $F_n(\bm{r})$, which represents the transition between the $n$th eigenstate $\phi_n$ of the quark channel and the scattering state of the hadron channel with the relative coordinate $\bm{r}$ as
\begin{align}
    \braket{\phi_n \mid V^t \mid \bm{r}_{h}}\equiv F_n(\bm{r}).
\end{align}

The first term on the right-hand side of Eq.~\eqref{eq_VDD} gives the coordinate representation of the potential originally operated in the hadron channel. The second term contains information of the quark degrees of freedom due to the effect of channel coupling, and is a sum of contributions from discrete eigenstates because the quark channel has a confinement potential. Since the numerator of the second term depends on the relative coordinates $\bm{r}$ and $\bm{r'}$ of the particles before and after the interaction independently, the matrix elements of $ V^{h}_\mathrm{ eff}(E)$ represent nonlocal potentials. It can be seen that, regardless of the properties of the transition potential $V^t$, the matrix elements of the effective potential between hadrons $ V^{h}_\mathrm{ eff}(E)$ are always nonlocal due to the channel coupling. Also, from the denominator of the second term, we find that the potential strength is energy dependent and diverges at the discrete eigenenergies in the quark channel $E=E_n$.

Next, we investigate the effect of the channel coupling in the inter-quark potential. The Schr\"{o}dinger equation for the hadron channel without the channel coupling is
\begin{align}
    (T^h +\Delta+ V^h)\ket{\psi} =E\ket{\psi},
\end{align}
with the wave function $\ket{\psi}$. $V^h$ is a scattering potential, so the Schr\"{o}dinger equation has continuous eigenstates $\ket{\bm{p}_\mathrm{full}}$ labeled by the momentum $\bm{p}$ as scattering solutions. Furthermore, when $V^h$ is sufficiently attractive, it can generate discrete eigenstates $\ket{\psi_n}$. These eigenstates satisfy the Schr\"{o}dinger equations
\begin{align}
    (T^h +\Delta+ V^h)\ket{\bm{p}_\mathrm{{full}}} & =E_{\bm{p}}\ket{\bm{p}_\mathrm{{full}}}, \\
    (T^h +\Delta+ V^h)\ket{\psi_n}                 & = E^h_n\ket{\psi_n}, \label{risan}
\end{align}
where $E_{\bm{p}}={\bm{p}^2}/({2m})+\Delta$ is the eigenenergy of a state with momentum $\bm{p}$ and $m$ is the reduced mass of the hadron channel. 
Since the energy of a state with zero momentum is the threshold energy, $E_{\boldsymbol{p}}=\Delta$ for $\boldsymbol{p}=\bm{0}$. Also, $E^h_n < \Delta$ is the eigenenergy of the $n$th state $\ket{\psi_n}$.
In this case, we define the normalization and orthogonality of eigenstates as
\begin{align}
    \braket{\psi_n \mid \psi_m}                                      & =\delta_{nm} ,             \\
    \braket{  \bm{p'}_\mathrm{{full}} \mid \bm{p}_\mathrm{{full}}  } & = \delta(\bm{p'}-\bm{p}) , \\
    \braket{\psi_n \mid \bm{p}_\mathrm{{full}} }                     & = 0.
\end{align}
Also, due to the completeness of the basis,
\begin{align}\label{eq_kanzen_h}
    \sum_n \ket{\psi_n}\bra{\psi_n} +\int d\bm{p} \ket{\bm{p}_\mathrm{{full}}}\bra{\bm{p}_\mathrm{{full}}}=1^h,
\end{align}
the identify operator $1^h$ of the hadron channel is expressed using $\ket{\bm{p}_\mathrm{full}}$ and $\ket{\psi_n}$.
The coordinate space potential of the quark channel is obtained as 
\begin{align}\label{eq_VCC}
     & \braket{\bm{r'}_{q} \mid V^{q}_\mathrm{ eff}(E)\mid\bm{r}_{q}} \nonumber                                                                                              \\
     & \quad =   \braket{\bm{r'}_{q} \mid V^{q}\mid\bm{r}_{q}}   +   \sum_n \frac{F^{h*}_n(\bm{r'}) F^h_n(\bm{r})}{E-E^h_n}
    \nonumber                                                                                                                                                                \\
     & \quad\quad + \int d\bm{p} \frac{\braket{\bm{r'}_{q} \mid V^t \mid \bm{p}_\mathrm{{full}}}\braket{\bm{p}_\mathrm{{full}}\mid V^t \mid \bm{r}_{q}}}{E-E_{\bm{p}}+i0^+},
\end{align}
where the complete set of the basis~\eqref{eq_kanzen_h} is used. We define the form factor $F^h_n(\bm{r})$, which stands for the transition between the $n$-th eigenstate $\psi_n$ of the hadron channel and the scattering state of the quark channel in relative coordinates $\bm{r}$ as
\begin{math}
    \braket{\psi_n \mid V^t \mid \bm{r}_{q}}\equiv F^h_n(\bm{r}).
\end{math}

The first term on the right-hand side of Eq.~\eqref{eq_VCC} expresses the coordinate representation of the potential inherent to the quark channel. The third term represents the effect of channel coupling, and is an integration of continuous eigenstates because the interaction in the hadron channel is a scattering potential. As with the effective interaction between hadrons in Eq.~\eqref{eq_VDD}, we find from the numerator of the integrand that the effective potential between quarks is a nonlocal potential. It can also be seen from the denominator that it is energy dependent.

Furthermore, the denominator of the third term shows that the effective potential has an imaginary part for $E \geq \Delta$ by picking up the pole in the $\abs{\bm{p}}$ integration :
\begin{align}\label{eq_ImVeff_q}
     & \mathrm{ Im} \left[\braket{\bm{r'}_{q} \mid V^{q}_\mathrm{ eff}(E)\mid\bm{r}_{q}}\right]\nonumber                                                           \\
     & \quad =
    4\pi^2  m \sqrt{2m(E-\Delta)} \nonumber                                                                                                                        \\
     & \quad \quad\times \braket{\bm{r'}_{q} \mid V^t \mid \bm{p}_\mathrm{{full}}}\braket{\bm{p}_\mathrm{{full}}\mid V^t \mid \bm{r}_{q}} \Theta(E-\Delta)       ,
\end{align}
Equation~\eqref{eq_ImVeff_q} shows the physical process in which the quark channel decays into a lower-energy scattering state of the hadron channel. In the case where the quark mass is infinitely heavy, the string breaking occurs in the static quark potential due to the coupling with the hadron channel~\cite{Bali:2005fu}, but the potential does not generate an imaginary part.
This is because, the mesons containing the static quark also has an infinitely heavy mass, and hence no scattering state is generated. 
On the other hand, we show that the effective potential between quarks with finite quark mass~\eqref{eq_VCC}, gives an imaginary part at energies higher than the threshold due to the coupling with scattering states with continuous eigenvalues.

\subsection{Local approximation}\label{sub_sc_local}
In the previous section, we have shown that the coupling between the quark channel and the hadron channel makes the effective potentials nonlocal. In this section, we investigate the physical picture of the interaction by approximating the nonlocal potential to a local one, because the physical properties of the nonlocal potentials are not clear. 
First, we briefly summarize the difference between the nonlocal and the local potentials with the energy dependence.
We define the state $\ket{\bm{r}}$ as an eigenstate of the relative coordinate operator $\hat{\bm{r}}$, which satisfies the orthonormal conditions
\begin{math}
    \braket{\bm{r'}\mid \bm{r}} = \delta(\bm{r'}-\bm{r})
\end{math}
and completeness
\begin{math}
    \int d^3 \bm{r} ~ \ket{\bm{r}}\bra{\bm{r}} = 1
\end{math}
.
By taking the matrix elements, the energy-dependent potential operator $V(E)$ in the coordinate representation is given by
\begin{align}\label{eq_nonlocalV}
    \braket{\bm{r'} \mid V(E) \mid \bm{r}} = V(\bm{r'},\bm{r},E).
\end{align}
We call this the nonlocal potential when the matrix elements depend on $\bm{r}\text{ and }\bm{r'}$ independently.
As an example of the nonlocal potentials, we show a separable potential
\begin{align}\label{eq_Vseparable}
    V(\bm{r},\bm{r'},E) =\omega(E) V(\bm{r})V(\bm{r'}),
\end{align}
which is a product of a function of $\bm{r}$ only and a function of $\bm{r'}$ only, with the potential strength $\omega(E)$.
In fact, the contribution from the channel coupling in the effective potentials~\eqref{eq_VDD} and \eqref{eq_VCC} are given in the separable form.
On the other hand, a potential that is proportional to the delta function $\delta(\boldsymbol{r'}-\boldsymbol{r})$ is called a local potential:
\begin{align}\label{eq_localV}
    \braket{\bm{r'} \mid V(E) \mid \bm{r}} = V(\boldsymbol{r},E)\delta(\boldsymbol{r'}-\boldsymbol{r}).
\end{align}
The local potentials have an advantage of having a clear physical picture.
For example, the interaction range of the potential can be seen from the $\bm{r}$ dependence.

The Schr\"{o}dinger equation with the energy-dependent nonlocal potential is
\begin{align}\label{eq_sch_eq_non-local}
    -\frac{1}{2m}\nabla^2 \psi(\bm{r}) + \int d^3\bm{r'} ~V(\boldsymbol{r},\boldsymbol{r'},E) \psi(\bm{r'}) & =E \psi(\bm{r}).
\end{align}
We see that the above equation contains differentiation, integration, and the energy dependence.
On the other hand, the Schr\"{o}dinger equation with the energy-dependent local potential~\eqref{eq_localV} is given by:
\begin{align}\label{eq_sch_Edep}
    -\frac{1}{2m}\nabla^2 \psi(\bm{r}) + V(\bm{r},E) \psi(\bm{r}) & =E \psi(\bm{r}),
\end{align}
which is a differential equation with energy dependence on both sides.
The Schr\"{o}dinger equation with the local potential~\eqref{eq_sch_Edep} has an advantage of having less computational cost than the equation with the nonlocal potential~\eqref{eq_sch_eq_non-local}, because the integration in the second term on the left-hand side can be analytically calculated.
In order to solve the Schr\"{o}dinger equation with energy dependence, it is necessary to first determine the potential $V(E)$ at a certain energy $E$, and then find the eigenenergy $E$ on the right-hand side. In general, the obtained eigenenergy $E$ does not always coincide with the energy $E$, which is used to determine the potential.
In order to obtain the solution, we need to solve the equation self-consistently so that the energies $E$ on both sides are consistent with each other.  

In the following, we introduce the formal derivative expansion and the derivative expansion by the HAL QCD method for potentials with energy dependence based on Ref.~\cite{Aoki:2021ahj}.
First, we formulate the formal derivative expansion, which is the method of approximating a nonlocal potential~\eqref{eq_Vseparable} to a local one by expanding it in powers of the derivative $\nabla$ by the Taylor expansion.
The Taylor expansion of the separable nonlocal potential around $\bm{r'}=\bm{r}$ is
\begin{align}\label{eq_formal_ippan}
    &\omega(E)V(\boldsymbol{r}) V(\bm{r'}) \nonumber \\
    &\quad = \sum_{n=0}^{\infty} V_n^{i_1\cdots i_n}(\bm{r},E)\nabla_{i_1}^{r} \cdots \nabla_{i_n}^{r} \delta^3 (\bm{r}-\bm{r'}),
\end{align}
where the expansion coefficients are
\begin{align}
     & V_n^{i_1\cdots i_n}(\bm{r},E) \nonumber                                                                                   \\
     & \quad  = \frac{1}{n!}\omega(E) V(\bm{r})\int d^3 \bm{r'}~V(\bm{r'}) (\bm{r'}-\bm{r})^{i_1} \cdots (\bm{r'}-\bm{r})^{i_n}.
\end{align}
Since $n$ is the order of derivative, the higher order terms of the expansion can be neglected for small momenta.
The most dominant contribution at low energies is given by the $n=0$ term:
\begin{align}\label{eq_formal_houhou}
    V_0(\bm{r},E) & = \omega(E) V(\bm{r})\int d^3 \bm{r'}~V(\bm{r'}).
\end{align}
Finally, we obtain the general form of the lowest order local potential that approximates the separable nonlocal potential~\eqref{eq_formal_ippan} by the formal derivative expansion as
\begin{align}
    V(\boldsymbol{r},\boldsymbol{r'},E)  = V_0(\bm{r},E)\delta(\bm{r'}-\bm{r}) +\mathcal{O}(\nabla^1).
\end{align}
We find that the local approximated potential by the formal derivative expansion $V_0(\bm{r},E)$ also has energy dependence when the nonlocal potential has energy dependence.

Next, we consider the derivative expansion by the HAL QCD method (hereafter, referred to as the HAL QCD method).
We first assume that the wave function of the nonlocal potential satisfies the Schr\"{o}dinger equation with a local potential.
We then solve inversely the Schr\"{o}dinger equation for the local potential with the given wave function of the nonlocal potential.  
In the following, we describe the procedure of the HAL QCD method for the energy-dependent nonlocal potential.
We consider the case where the wave function $\psi_k(\bm{r})$ at energy $E$ has already been obtained for a certain nonlocal potential where the eigenmomentum is $k=\sqrt{2mE}$.
First, we prepare wave functions $\psi_{k_i}(\bm{r})$ at $n+1$ points of eigenmomenta $k_i,\,(i=0,1,\cdots , n)$. We assume that the wave functions satisfy the Schr\"{o}dinger equation with the local potential $V_n(\bm{r},\bm{\nabla})$ without energy dependence:
\begin{align}\label{eq_sch_HAL}
    \left(-\frac{1}{2m}\bm{{\nabla}}^2+V_n(\bm{r},\bm{\nabla})\right)\psi_{k_i}(\bm{r}) & =E_{k_i}\psi_{k_i}(\bm{r}),
\end{align}
where $E_{k_i}$ is the energy $E_{k_i}={k_i^2}/{2m}$ at the momentum $k_i$.
Furthermore, the expansion of the potential $V_n(\bm{r},\bm{\nabla})$ in powers of the derivative with $n+1$ terms leads to
\begin{align}\label{eq_VHAL_define}
    V_n(\bm{r},\bm{\nabla})=\sum_{i=0}^n V_{n,i}(\bm{r})(\bm{\nabla}^2)^i .
\end{align}
The $n+1$ expansion coefficients $V_{n,i}(\bm{r})$ are determined by the $n+1$ Schr\"{o}dinger equations~\eqref{eq_sch_HAL}. Here, we adopt the formulation with only even powers of derivatives following Ref.~\cite{Aoki:2021ahj}.
We find from Eq.~\eqref{eq_sch_HAL} that the expansion coefficients of the potential $V_{n,i}(\bm{r})$ satisfy
\begin{align}\label{eq_VHAL_exact}
    \sum_{j=0}^n T_{ij}(\bm{r} )V_{n,j}(\bm{r})&= K_i(\bm{r}), \\
    T_{ij}(\bm{r}) & \equiv {\nabla^{2j}}\psi_{k_i}(\bm{r}),                             \\
    K_i(\bm{r})    & \equiv \frac{1}{2m}\left( k_i^2+\nabla^2 \right)\psi_{k_i}(\bm{r}).
\end{align}
Therefore, $V_{n,i}(\bm{r})$ is obtained as
\begin{align}\label{eq_V_tenkaikeisuu}
    V_{n,i}(\bm{r}) = \sum_{j=0}^n \left[ T^{-1}(\bm{r})\right]_{ij}K_j(\bm{r}).
\end{align}
The approximated local potential~\eqref{eq_VHAL_define} obtained by the HAL QCD method is an energy-independent potential even when the nonlocal potential has energy dependence. This is because, when the potential $V_n(\bm{r},\bm{\nabla})$ includes energy $E_{k_i}$ dependence, $(n+1)^2$ simultaneous equations are required to determine the expansion coefficients of the potential. However, we can only have $n+1$ Schr\"{o}dinger equations~\eqref{eq_sch_HAL}, so we cannot solve all the expansion coefficients.
It should be noted that the potential $V_n$ in Eq.~\eqref{eq_VHAL_define} is an approximated potential that depends on the selection of $k_i\,(i=0,1,2,\cdots , n)$. 
Although the approximated potential is guaranteed to give the correct phase shift at the points $k=k_i\,(i=0,1,2,\cdots , n)$, it provides approximated values at the other momenta.

At the lowest order of derivatives ($n=0$), we can only take $i=j=0$ in Eq.~\eqref{eq_V_tenkaikeisuu}, so the local approximated potential is given only by the wave function at a single momentum $k_0$ as
\begin{align}\label{eq_VHAL_wf}
    V_0(\bm{r},\nabla;k_0) =V_{0,0}(\bm{r};k_0)
    = \frac{k_0^2}{2m}+ \frac{\nabla^2\psi_{k_0}(\bm{r})}{2m\psi_{k_0}(\bm{r})}.
\end{align}
Thus, we obtain the lowest order of local approximated potential by the HAL QCD method as
\begin{align}\label{eq_VHAL_kaiseki}
     & V_0(\bm{r},\bm{r}',E)  =V_{0,0}(\bm{r};k_0)\delta(\bm{r}'-\bm{r})+\mathcal{O}(\nabla^2).
\end{align}
The error of the approximation is of the second order or higher because the local approximated potential by the HAL QCD method~\eqref{eq_VHAL_define} is defined not to include the odd powers of derivatives.
It is notable that although the local potential by the HAL QCD method is energy independent, the potential depends on the momentum $k_0$ that specifies the wave function $\psi_{k_0}$ used to determine the potential $V_0(\bm{r},\bm{r}',E) $. The Schr\"{o}dinger equation with the local potential obtained by the HAL QCD method~\eqref{eq_VHAL_kaiseki} is given by
\begin{align}\label{eq_sch_V00}
    \left[-\frac{1}{2m}\nabla^2 +V_{0,0}(\bm{r};k_0) \right]\psi(\bm{r}) & =E\psi(\bm{r}).
\end{align}
When we solve this equation at the energy $E = k_0^2/(2m)$, we obtain the wave function $\psi=\psi_{k_0}$ and the phase shift is equivalent to that obtained from the nonlocal potential.
The exact wave functions, however, is not always obtained when we solve the Schr\"{o}dinger equation~\eqref{eq_sch_V00} at $E\neq {k_0}^2/(2m) $ in general.

\subsection{Analytic form with Yukawa form factor}\label{sub_sc_yukawa}
In order to analytically discuss the properties of the local approximated potentials, we consider the separable nonlocal potential~\eqref{eq_Vseparable} with the Yukawa-type function as the form factor
\begin{align}
    V(\bm{r})&=\frac{e^{-\mu r}}{r}, \label{eq_Vr_yukawa} \\
    V(\boldsymbol{r},\boldsymbol{r'},E)&= \omega(E) V(\bm{r})V(\bm{r'}) = \omega(E) \frac{e^{-\mu r}}{r}\frac{e^{-\mu r'}}{r'},\label{eq_Vyukawa}
\end{align}
where $r=\abs{\bm{r}}$, $\mu$ is a cutoff constant, and $\omega(E)$ is an energy-dependent coefficient that determines the strength of the interaction, where a positive sign indicates the repulsion and a negative sign the attraction.
The nonlocal potential with the Yukawa-type form factor~\eqref{eq_Vyukawa} is a special example of nonlocal potentials that can be solved analytically. In fact, in Ref.~\cite{Aoki:2021ahj}, the local potential was analytically obtained by two methods, the formal derivative expansion and the HAL QCD method, for the energy-independent potential strength $\omega$.

We can obtain the analytical form of the local approximated potential in the formal derivative expansion, by substituting Eq.~\eqref{eq_Vr_yukawa} into Eq.~\eqref{eq_formal_houhou}, and the approximation at the lowest order of derivatives gives
\begin{align}
    V_\mathrm(\bm{r},\bm{r'},E) & = V^\mathrm{{formal}}(r,E)\delta(\bm{r}-\bm{r'})+\mathcal{O}(\nabla), \nonumber \\
    V^\mathrm{{formal}}(r,E)                & =\omega(E)\frac{4\pi}{\mu^2}\frac{e^{-\mu r}}{r}.\label{eq_Vformalyukawa}
\end{align}
Equation~\eqref{eq_Vformalyukawa} is consistent with the result given in Ref.~\cite{Aoki:2021ahj}, where the energy dependence of $\omega$ is simply included. We find that the local potential by the formal derivative expansion has the same energy dependence as the original nonlocal potential. Therefore, we need to solve the Schr\"{o}dinger equation~\eqref{eq_sch_Edep} self-consistently for the energy dependence. Also, focusing on the $r$ dependence, we see that the range of the potential is $1/\mu$, just like the properties of the Yukawa-type potential.
The cutoff $\mu$ is the momentum that characterizes the form factor of the transition potential caused by the channel coupling. It is noteworthy that the range of the hadron-hadron potential is determined also by $\mu$.

Next, we first obtain the scattering wave function and phase shift of the nonlocal potential for the application of the HAL QCD method. By solving the Lippmann-Schwinger equation for the positive a energy $E$, as in Ref.~\cite{Aoki:2021ahj} for the energy-independent potential, we obtain the scattering wave function analytically as
\begin{align}\label{eq_psi_yukawa}
    \psi_{k}(r) = \frac{\sin[kr+\delta(k)]-\sin\delta(k)e^{-\mu r}}{kr},
\end{align}
where the magnitude of momentum is $k=\sqrt{2mE}$, and the scattering phase shift $\delta(k)$ is given as
\begin{align}\label{eq_kcotdelta}
    k \cot\delta(k)
     & = -\frac{\mu[4\pi m \omega(E)+\mu^3]}{8\pi m \omega(E)}                                                \\ \nonumber
     & \quad +\frac{1}{2\mu}\left[1-\frac{2\mu^3}{4\pi m \omega(E)}\right]k^2-\frac{1}{8\pi m \omega(E)}k^4 .
\end{align}
Furthermore, from Eq.~\eqref{eq_kcotdelta}, the scattering length $a_0$ is obtained as
\begin{align}\label{eq_a0}
    a_0=+\frac{8\pi m \omega(E=0)}{\mu[4\pi m \omega(E=0)+\mu^3]}.
\end{align}
We find that the analytical solution can be obtained for the energy-dependent potential by replacing $\omega \to \omega(E)$ in the expressions for energy-independent potentials. 
This is because the energy is fixed when we solve for the scattering wave function, so that the self-consistency of the energy is not necessarily imposed in the Lippmann-Schwinger equation.
On the other hand, it is necessary to solve self-consistently the Schr\"{o}dinger equation for the bound state.
Therefore, it should be noted that the wave function of the bound state cannot be obtained by the simple replacement of $\omega \to \omega(E)$.

We next obtain the local approximation by the HAL QCD method analytically at the lowest order, by substituting the wave function~\eqref{eq_psi_yukawa} into Eq.~\eqref{eq_VHAL_kaiseki} as
\begin{align}\label{eq_VHAL_gutai}
     & V^\mathrm{ HAL}(r;k_0) = \frac{k_0^2}{2m}  \nonumber                                                                                                           \\
     & \quad   +  \frac{-k_0^2\sin \left[k_0 r+\delta(k_0)\right]-\mu^2\sin\delta(k_0) e^{-\mu r}}{2m\{\sin\left[k_0 r+\delta(k_0)\right]-\sin\delta(k_0) e^{-\mu r}\}}.
\end{align}
Here, the phase shift $\delta(k_0)$ is given by Eq.~\eqref{eq_kcotdelta}. In this way, we derive an explicit expression for the local approximated potential by the HAL QCD method.

To investigate the low-energy scattering, we show the $k_0 \to 0$ limit of the local approximated potential by the HAL QCD method~\eqref{eq_VHAL_gutai} with the scattering length $a_0$ as 
\begin{align}\label{eq_VHALE0gutai}
    V^{\mathrm{ HAL}}(r;k_0=0)
    = \frac{a_0\mu^2e^{-\mu r}}{2m\left( r-a_0+a_0e^{-\mu r} \right)}.
\end{align}
Similarly, we obtain the wave function~\eqref{eq_psi_yukawa} in the ${k_0 \to 0}$ limit with the scattering length $a_0$ as
\begin{align}\label{eq_VHAL_zerolimit}
    \lim_{k_0 \to 0}\psi_{k_0}(r)
    =  \frac{r-a_0+a_0e^{-\mu r}}{r}.
\end{align}
Here, we have used the relation that
\begin{math}
    \lim_{k \to 0} \sin\delta(k)=-ka_0.
\end{math}
\section{Application to $X(3872)$}\label{sc_Application}
\subsection{Model for $X(3872)$}\label{sub_sc_form_X}
In this section, we first construct a model of $X(3872)$ to apply the formulation of Sec.~\ref*{sc_formu}. The most plausible possibility of the structure of the exotic hadron $X(3872)$ is a mixture of the $c\bar{c}$ quarkonium state with the quark degrees of freedom and the $D\bar{D}^{*}$ molecular component with the hadron degrees of freedom~\cite{Takizawa:2012hy,Hosaka:2016pey}. In this work, we consider only the coupling between the neutral channel ($D^{0}\bar{D}^{*0}$) having the closest threshold to $X(3872)$ and the $c\bar{c}$ state to focus on the effect of the channel coupling to the quark degrees of freedom in the hadron-hadron potential.
We do not consider the coupling to the charged channels ($ D^\pm D^{*\mp}$), decay channels ($J\psi \pi \pi$, etc.), and the direct interactions between the heavy mesons for simplicity.

We apply the formulation of the channel coupling in Sec.~\ref{sub_sc_channel} to $X(3872)$. Here, we take $c\bar{c}$ as the quark channel $\ket{q}$ and $D^0\bar{D}^{*0}$ as the hadron channel $\ket{h}$. To achieve $J^{PC}=1^{++}$ of $X(3872)$, the $D^0\bar{D}^{*0}$ channel is combined in the $s$ wave\footnote{Hereafter, ``$D^0\bar{D}^{*0}$'' stands for the abbreviation of the linear combination $(\ket{D^0\bar{D}^{*0}}+\ket{D^{*0}\bar{D}^{0}})/\sqrt{2}$ with the charge conjugation $C=+$.} and the $c\bar{c}$ channel is combined into the $^3{\rm P}_1$ state. 
Equation~\eqref{eq_2ch_hamiltonian} gives the Hamiltonian $H$ for these channels as
\begin{align}\label{eq_2ch_hamiltonian_X}
    H=
    \begin{pmatrix}
        T^{c\bar{c}} & 0                           \\
        0            & T^{D^0\bar{D}^{*0}}+ \Delta
    \end{pmatrix}
    +
    \begin{pmatrix}
        V^{c\bar{c}} & V^t \\
        V^t          & 0
    \end{pmatrix}.
\end{align}
For the description of the dynamics in the quark channel $T^{c\bar{c}}+V^{c\bar{c}}$, we consider constituent quark model. In a standard constituent quark model with the Cornell potential in the $c\bar{c}$ channel,
one obtains the $\chi_{c1}(1P)$ charmonium as the ground state, and the $\chi_{c1}(2P)$ state at slightly higher than the $D^0\bar{D}^{*0}$ threshold energy as the first excited state.
Hereafter, we denote the $\chi_{c1}(2P)$ state, which has the strongest effect on the $D^0\bar{D}^{*0}$ channel, as $\phi_0$ and consider only $\ket{\phi_0}$ among the eigenstates in the $\ket{q}$ channel. Denoting the mass of $\phi_0$ as $m_{\phi_0}$, we define the energy relative to the threshold of the $D^0\bar{D}^{*0}$ channel as $E_0$: 
\begin{align}
    E_0=m_{\phi_0}-(m_{\bar{D}^{*0}} +m_{D^0}),
\end{align}
where $m_{\bar{D}^{*0}} $ and $m_{D^0}$ are the masses of ${\bar{D}^{*0}} $ and ${D^0}$, respectively.

Next, we define the matrix element of the transition potential $V^t$ between $\phi_0$ and $D^0\bar{D}^{*0}$ in the coordinate representation to be of Yukawa type as
\begin{align}
    \braket{\bm{r}_{D^0\bar{D}^{*0}}|V^t|\phi_0}=g_0\frac{e^{-\mu r}}{r},
\end{align}
where $\mu$ is a cutoff, which is a parameter that gives the range of the interaction as $1/\mu$. The coupling constant $g_{0}$ is determined in this study to reproduce the mass of $X(3872)$ as
\begin{align}\label{eq_g02_define}
    g_0^2 & = \frac{m_{\phi_0}-m_{X(3872)}}{I},                                                                                     \\
    I     & =\int_0^{\infty}\frac{ dk ~8 k^2 }{m_{D^0}+m_{D*^{0}}+\frac{k^2}{2m}-m_{X(3872)}} \left( \frac{1}{k^2+\mu^2}\right) ^2,
\end{align}
where $m_{X(3872)}$ is the mass of $X(3872)$ and $m$ is the reduced mass of the ${\bar{D}^{*0}} $ and ${D^0}$ system.

In this setup, we obtain the nonlocal effective $D^0\bar{D}^{*0}$ potential coupled with the $\phi_{0}$ state by applying the hadron-hadron effective potential~\eqref{eq_VDD} developed in Sec.~\ref*{sc_formu} to the model of $X(3872)$ as
\begin{align}\label{eq_VDDeff}
    V_\mathrm{ eff}^{D^0\bar{D}^{*0}}(\boldsymbol{r},\boldsymbol{r'},E)   = \frac{g_0^2}{E-E_0}\frac{e^{-\mu r}}{r}\frac{e^{-\mu r'}}{r'}.
\end{align}
We find that this potential is classified as the separable and the Yukawa-type potential by identifying the potential strength as
\begin{align}\label{eq_omega_X}
    \omega(E)=\frac{g_0^2}{E-E_0},
\end{align}
in Eq.~\eqref{eq_Vyukawa}.
The local potential by the formal derivative expansion can be obtained from the nonlocal effective potential of $X(3872)$ in Eq.~\eqref{eq_VDDeff}, by substituting the potential strength $\omega(E)$ of Eq.~\eqref{eq_omega_X} into the general form of Eq.~\eqref{eq_Vformalyukawa}.

The local potential by the HAL QCD method~\eqref{eq_VHAL_gutai} and its zero-energy limit~\eqref{eq_VHALE0gutai} are also obtained for the nonlocal effective potential of $X(3872)$ in the same expressions. Here, substituting Eq.~\eqref{eq_omega_X} into Eq.~\eqref{eq_kcotdelta} we obtain the explicit expression of the phase shift $\delta(k_0)$, which is given up to the sixth order of $k_0$ as 
\begin{align}\label{eq_exact_delta}
     & k_0 \cot\delta(k_0) \nonumber                                                                                                                                            \\
     & = -\frac{\mu}{2}\left[1-\frac{\mu^3 E_0}{4\pi m g_0^2}\right]   +\frac{1}{2\mu}\left[1-\frac{\mu^5}{8\pi m^2 g_0^2}+\frac{\mu^3 E_0}{2\pi mg_0^2}\right]k_0^2  \nonumber \\
     & \quad  +\left[-\frac{\mu^2}{8\pi m^2 g_0^2}+\frac{E_0}{8\pi m g_0^2}\right]k_0^4-\frac{1}{16\pi m^2 g_0^2}k_0^6,
\end{align}
The scattering length $a_0$ is determined from the constant term in Eq.~\eqref{eq_exact_delta} as
\begin{align}\label{eq_a0_gutai}
    a_0 & = \frac{8\pi m g_0^2}{\mu(4\pi m g_0^2-\mu^3E_0)}.
\end{align}
\subsection{Numerical results}\label{sub_sc_nmrcl_result}
In this section, we study the properties of the local approximated effective $D^{0}\bar{D}^{*0}$ potentials for $X(3872)$ by the numerical calculations. We also compare the scattering phase shifts calculated from the local potentials with the exact scattering phase shift calculated from the original nonlocal potential.
The hadron masses are taken from the central values in PDG~\cite{ParticleDataGroup:2022pth}, which gives the binding energy $B_{X(3872)}$ of $X(3872)$ as $B_{X(3872)}=m_{D^0}+m_{\bar{D}^{*0}}-m_{X(3872)}\simeq 40$ keV. Namely, $X(3872)$ is a very shallow bound state.
Since the bare mass of the $\chi_{c1}(2P)$ state $m_{\phi_0}$ is not an observable quantity, we determine it to be $3.950~\mathrm{GeV}$ based on the quark model in Ref.~\cite{Godfrey:1985xj}. In this case, we obtain $E_0 \simeq0.078$ GeV and the bare state appears above the threshold.
The cutoff $\mu$ is chosen as the mass of $\pi$, which is the lightest exchange meson between $D^0\bar{D}^{*0}$, or as the mass of $\rho$ for the description of the shorter range potential.
Since there are no analytic solutions of the phase shift both from the local potentials obtained by the formal derivative expansion~\eqref{eq_Vformalyukawa} and by the HAL QCD method~\eqref{eq_VHAL_gutai}, we evaluate the phase shift by the numerical calculation.

\begin{figure}[tbp]
    \centering
    \includegraphics[bb=0 0 1791 1303,width=0.48\textwidth]{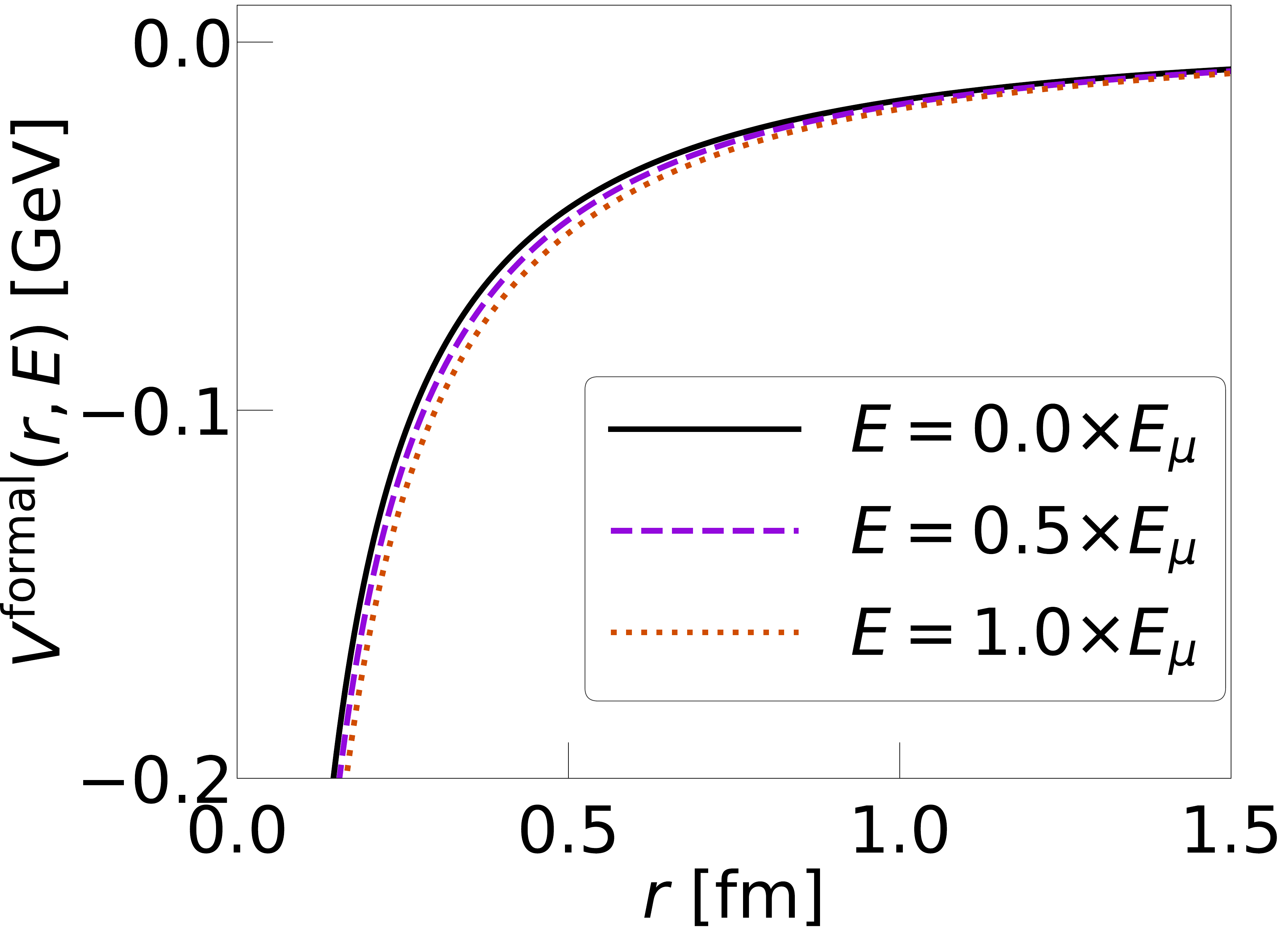}
    \caption{The local potentials by the formal derivative expansion~$V^\mathrm{{formal}}(r,E)$ as functions of the relative distance $r$ with $E=0$ (solid line), $E=E_{\mu}/2$ (dashed line), and $E=E_{\mu}$ (dotted line).}
    \label{fig_Vformal}
\end{figure}

\begin{figure}[tbp]
    \centering
    \includegraphics[bb=0 0 1848 1342,width=0.48\textwidth]{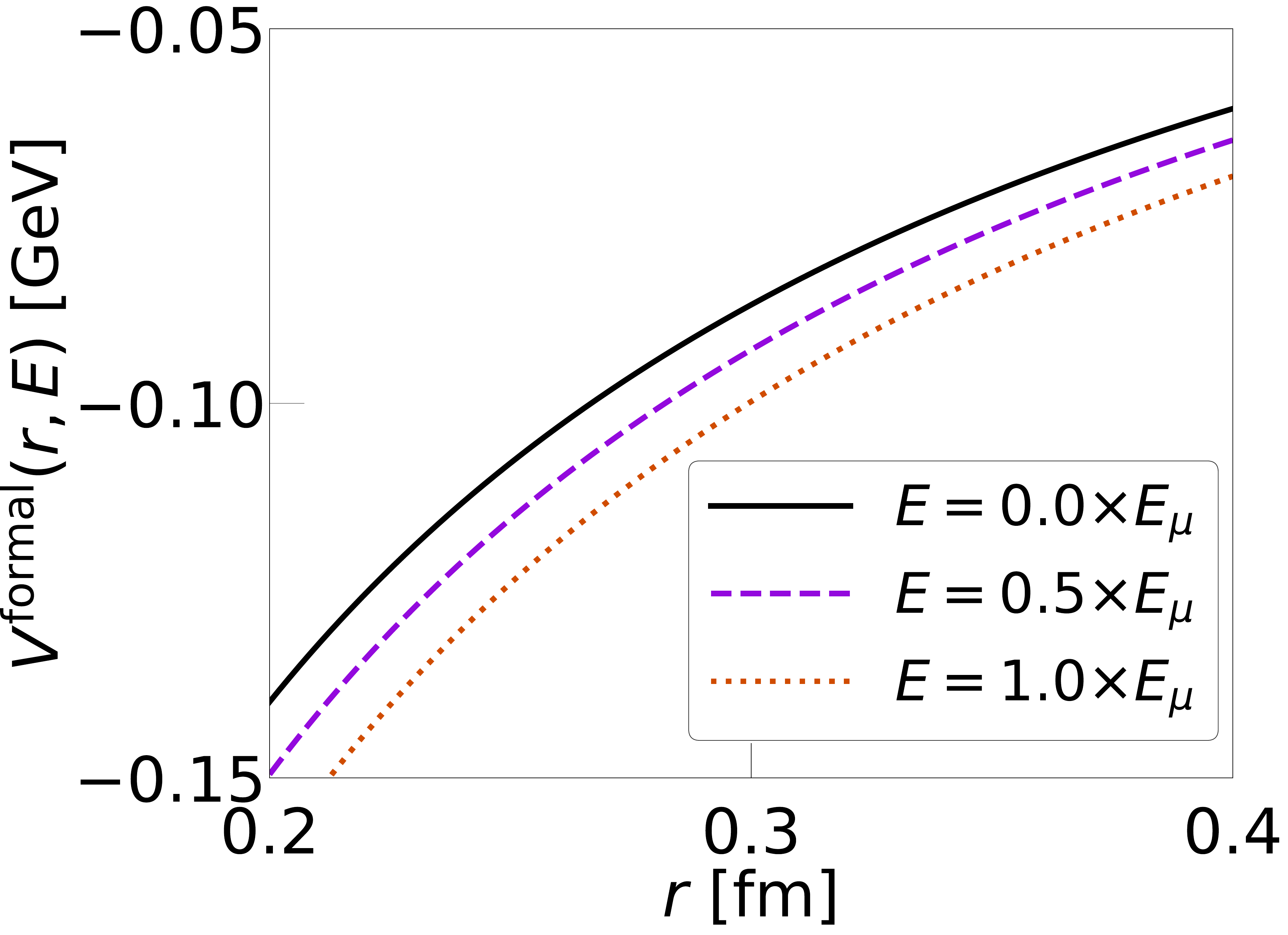}
    \caption{The enlargement of Fig.~\ref{fig_Vformal} around 
    $r=0.3 ~\mathrm{ fm}$.
    }
    \label{fig_Vformalclose}
\end{figure}

Figure~\ref{fig_Vformal} shows the local $D^0\bar{D}^{*0}$ potentials by the formal derivative expansion~\eqref{eq_Vformalyukawa} as functions of the relative distance $r$ with $\mu=0.14$ GeV. Its enlargement around $r=0.3 ~\mathrm{ fm}$ is shown in Fig.~\ref{fig_Vformalclose}. We set the upper bound of the energy in this model $E_{\mu}$ as that corresponding to the momentum cutoff $\mu$ as
\begin{align}
    E_{\mu}=\frac{\mu^2}{2m}.
\end{align}
To investigate the energy dependence of the potential, we show $V^\mathrm{{formal}}(r,E)$ with three different values, $E=0,E_{\mu}/2$, and $E_{\mu}$.
We see from Fig.~\ref*{fig_Vformal} that the potentials are attractive because the potential strength in Eq.~\eqref{eq_omega_X} is negative for $E\leq E_0$ as a consequence of the second-order perturbations, which act attractively on the ground state in quantum mechanics.
It is also seen that the local potentials show the Yukawa-type $r$ dependence as indicated in Eq.~\eqref{eq_Vformalyukawa}.

We find from Fig.~\ref*{fig_Vformal} that the energy dependence of the potentials~\eqref{eq_Vformalyukawa} with~\eqref{eq_omega_X} on energy $E$ is not very strong.
This is because the energy of the potential is chosen for $E=E_{\mu}\simeq 0.01 ~\mathrm{ GeV}$ at the maximum, which is much smaller than the bare energy $E_0 \simeq 0.078 ~\mathrm{ GeV}$. In this case, the potential strength~\eqref{eq_omega_X} shows a weak energy dependence in the range of $0\leq E\leq E_{\mu}$. 
The local potential would show a strong energy dependence near $E \simeq E_0$ if a larger cutoff $\mu$ were chosen such that $E_{0}<E_{\mu}$.
Although the energy dependence of the local potential with $\mu=0.14$ GeV is small, the enlarged Fig.~\ref*{fig_Vformalclose} shows that the potential becomes more attractive as the energy $E$ increases.
This is because the factor of $E-E_0$ in the denominator of the potential strength~\eqref{eq_omega_X} enlarges the attraction as the energy $E$ increases, in the range of $E \leq E_0$.

\begin{figure}[tbp]
    \centering
    \includegraphics[bb=0 0 1793 1337, width=0.48\textwidth]{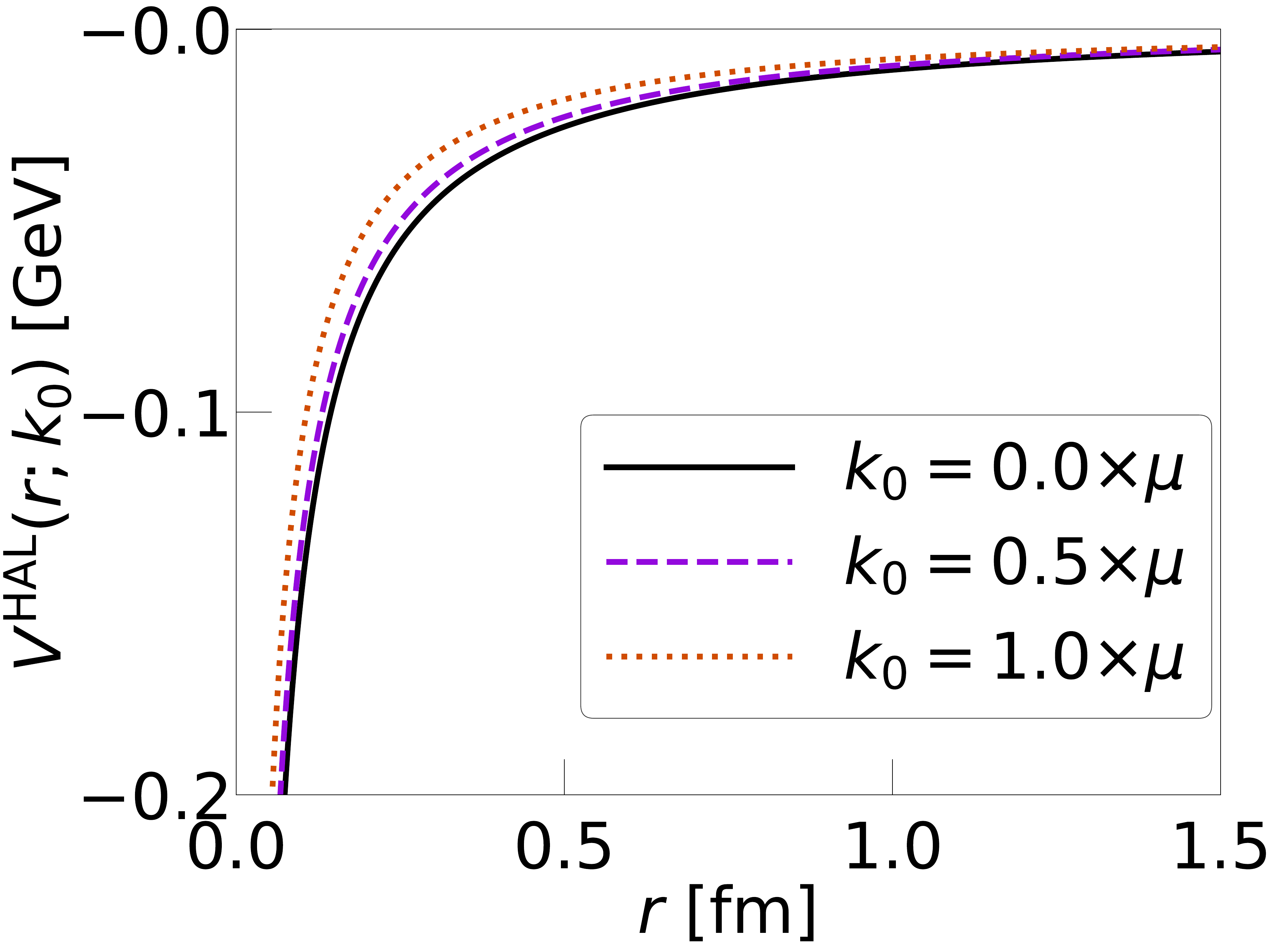}
    \caption{The local potentials by the HAL QCD method~$V^\mathrm{{HAL}}(r;k_0)$ as functions of the relative distance $r$ with $k_0=0$ (solid line), $k_0=\mu/2$ (dashed line), and $k_0=\mu$ (dotted line).}
    \label{fig_VHAL}
\end{figure}

\begin{figure}[tbp]
    \centering
    \includegraphics[bb=0 0 1850 1342, width=0.48\textwidth]{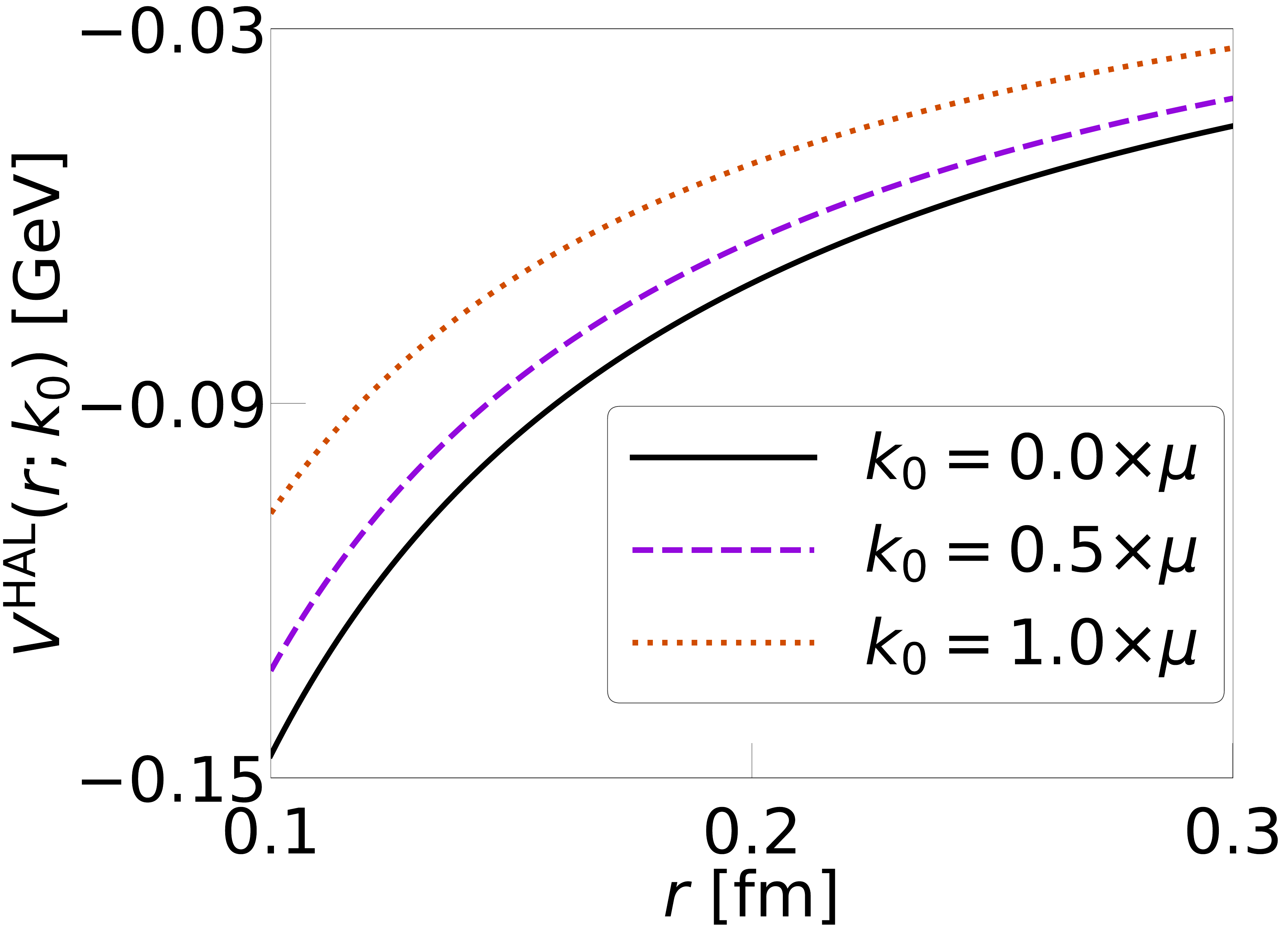}
    \caption{The enlargement of Fig.~\ref{fig_VHAL} around 
    $r=0.2 ~\mathrm{ fm}$.
    }
    \label{fig_VHALclose}
\end{figure}

We next show the local potentials by the HAL QCD method~\eqref{eq_VHAL_gutai} in Fig.~\ref{fig_VHAL} and its enlargement around $r=0.2 ~\mathrm{ fm}$ in Fig.~\ref{fig_VHALclose}.
The HAL QCD method yields an energy-independent potential unlike the formal derivative expansion discussed above. However, the potential depends on the momentum $k_0$ that specifies the wave function in constructing the potential, as seen in Eq.~\eqref{eq_VHAL_wf}. Therefore,
we show three different potentials with $k_0 = 0, \mu/2, \mu$ to investigate the $k_0$ dependence.
We find from Fig.~\ref*{fig_VHAL} that the potentials by the HAL QCD method~\eqref{eq_VHAL_gutai} do not strongly depend on the momentum $k_0$. The enlarged Fig.~\ref*{fig_VHALclose} indicates that the attractive potential strength decreases, as $k_0$ increases.

In both the approximation methods, there exists a set of parameters with which the potential strength diverges.
The potential by the formal derivative expansion~\eqref{eq_Vformalyukawa} diverges at $E=E_0$ due to the $E-E_0$ factor in the denominator. There are, however, no signs of the divergence in Fig.~\ref*{fig_Vformal} because the chosen energy is sufficiently smaller than the bare energy, $E \ll E_0$.
Because the potential by the HAL QCD method~\eqref{eq_VHAL_wf} contains the wave function in the denominator, it diverges at the nodes of the wave function [$\psi_{k_{0}}(r)=0$]~\cite{Nochi:2016wqg}. The first divergence of the potential occurs at $r \simeq 4.5~\mathrm{fm}$ and $3.8~\mathrm{fm}$ for $k_0^2/(2m) =E_{\mu}/2$ and $E_{\mu}$, respectively.
As we increase $k_{0}$, the potential diverges at smaller $r$ as the nodes of the wave function move toward the origin due to the rapid oscillations of the wave function for the larger wave number $k_0$. Therefore, one should cautious about the choice of the momentum $k_{0}$  in the HAL QCD method in order to avoid the pathological behavior of the potential in the small $r$ region.

\begin{figure}[tbp]
    \centering
    \includegraphics[bb=0 0 1708 1257,width=0.48\textwidth]{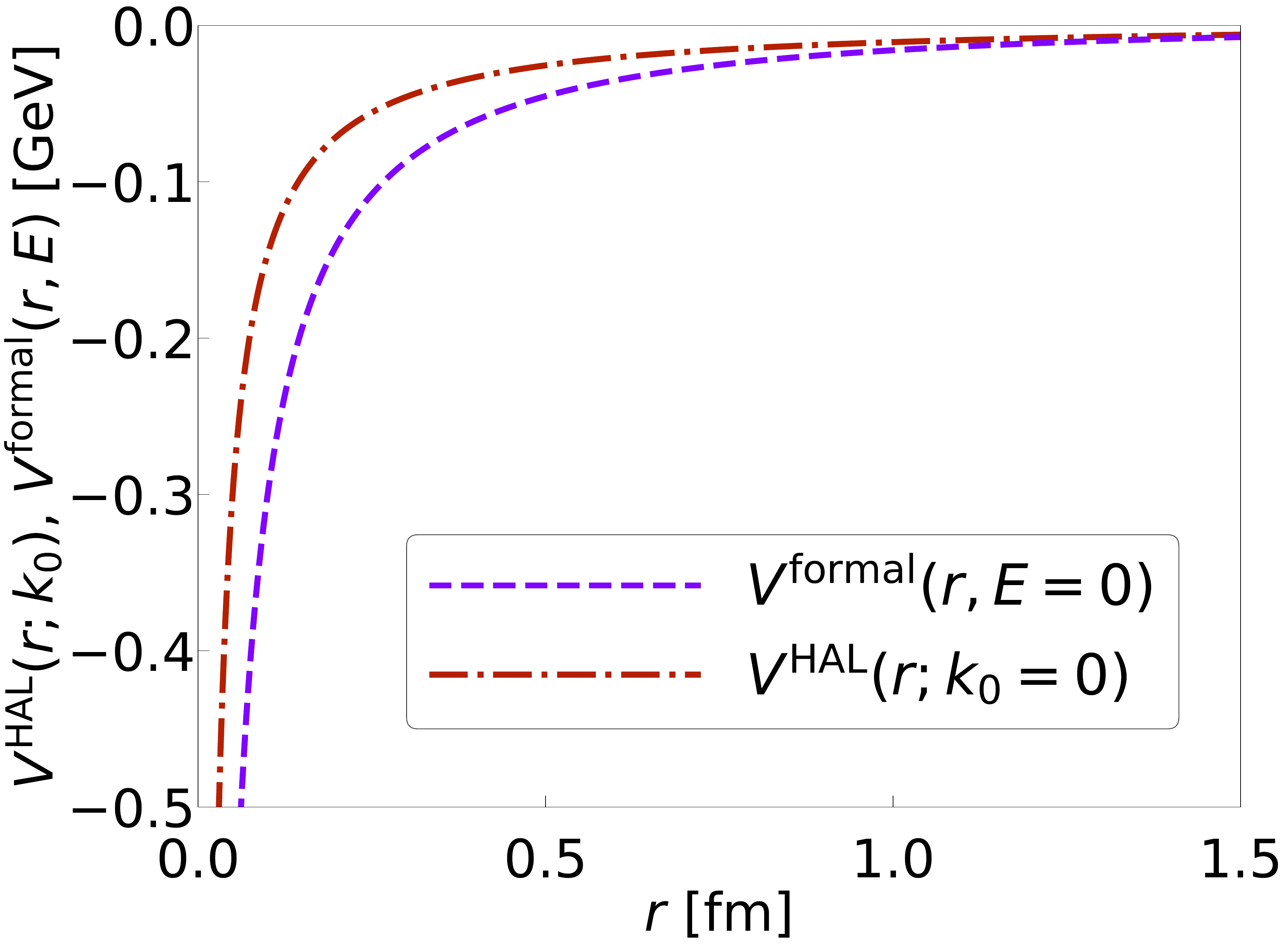}
    \caption{The comparison of the local potentials by the formal derivative expansion $V^\mathrm{{formal}}(r,E=0)$ (dashed line) and by the HAL QCD method $V^{\mathrm{ HAL}}(r;k_0=0)$ (dasehed-dotted line) as functions of the relative distance $r$.}
    \label{fig_Vcompare}
\end{figure}

Furthermore, we compare the local potential by the formal derivative expansion with $E=0$ (dashed line) and the one by the HAL QCD method with $k_0=0$ (dashed-dotted line) in Fig.~\ref{fig_Vcompare}.
We find the quantitative difference of the potentials, even though both are constructed from the same original nonlocal potential. In particular, the difference is pronounced in the short distance region of $r\lesssim 0.4 ~\mathrm{ fm}$. From the results in Figs.~~\ref*{fig_Vformal} and \ref*{fig_VHAL}, we find that the difference between the two local potentials is larger than the deviation due to the choice of the energy $E$ and the momentum $k_0$.

\begin{figure}[tbp]
    \centering
    \includegraphics[bb=0 0 1706 1250,width=0.45\textwidth]{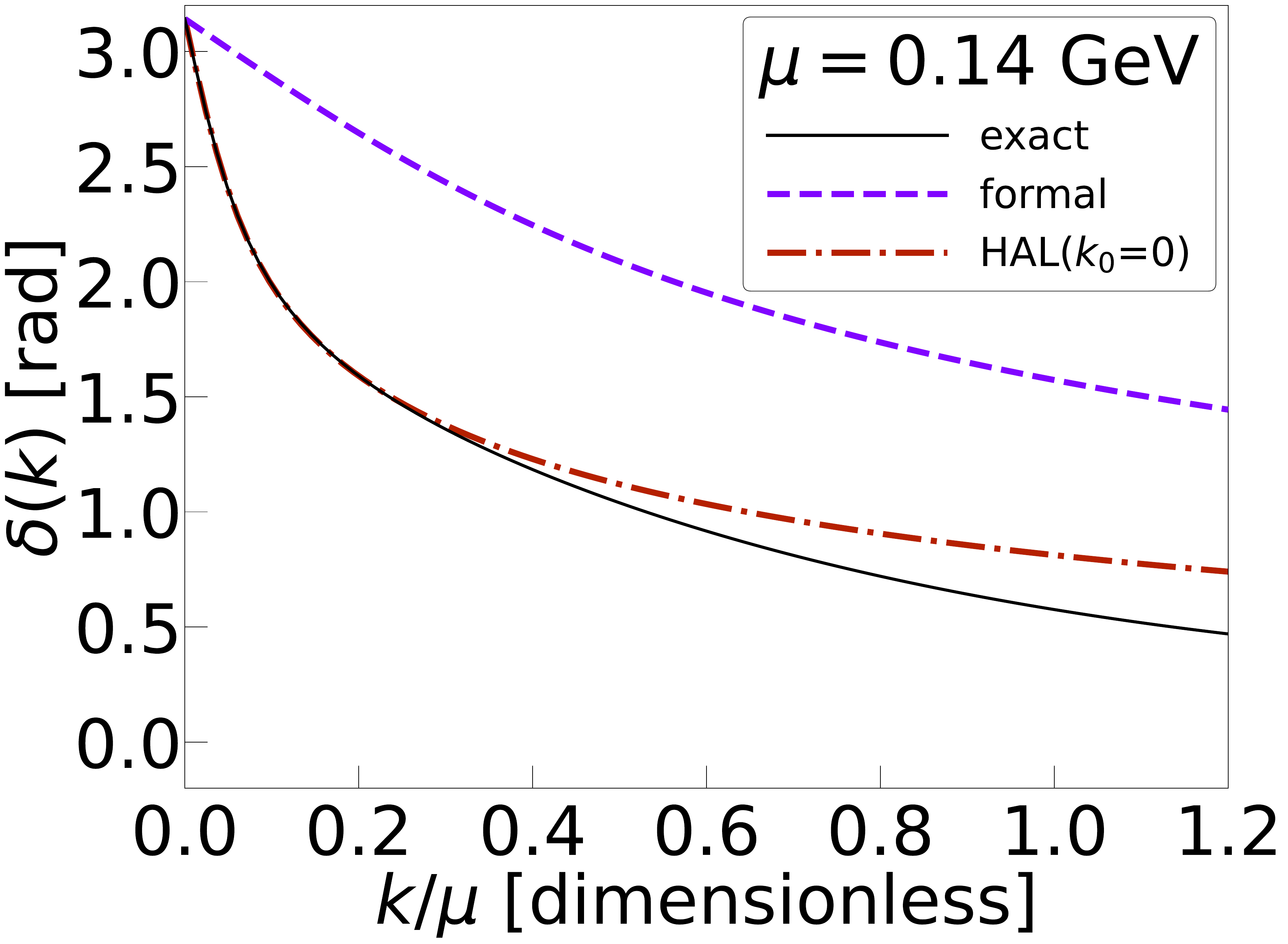}
    \caption{
    The phase shifts from the local potentials by the formal derivative expansion (dashed line) and by the HAL QCD method (dashed-dotted line) as functions of the dimensionless momentum $k/\mu$ with the cutoff $\mu=0.14$ GeV in comparison with the exact phase shift (solid line).
    }
    \label{fig_comparedelta}
\end{figure}

We next investigate how the difference of the local potentials affect on physical quantities by numerical calculation of the phase shift $\delta(k)$.
We calculate the phase shit at the momentum $k$ by varying the energy $E=k^2/(2m)$ in the potential by the formal derivative expansion. In contrast, the energy-independent potential by the HAL QCD method is treated in a standard for a given momentum $k_{0}$.\footnote{If we use the potential with $k_{0}=k$ for the calculation of the phase shift at $k$ as in the energy-dependent case, the exact phase shift is obtained for any $k$. Here we consider the fixed-$k_{0}$ potential to examine the accuracy of the local approximation.}
We show the phase shifts $\delta$ by the formal derivative expansion (dashed line) and by the HAL QCD method (dashed-dotted line) as functions of the dimensionless momentum $k/\mu$ with the cutoff $\mu$ in Fig.~\ref{fig_comparedelta}.
By comparing the dashed line with the dotted line, we find that the difference in potentials seen in Fig.~\ref*{fig_Vcompare} affects the phase shifts quantitatively.

We also show the exact phase shift $\delta$ (solid line) by the original nonlocal potential~\eqref{eq_exact_delta} in Fig.~\ref*{fig_comparedelta}.
We find that the exact phase shift is better approximated by the HAL QCD method than the formal derivative expansion.
In particular, the potential by the HAL QCD method works well in the small $k$ region, indicating that the scattering length defined by the slope of the phase shift at $k=0$ is also reproduced. In Table~\ref*{tab_a0_compare}, we summarize the exact scattering length~\eqref{eq_a0_gutai} and those by the local potentials numerically calculated by $\lim_{k \to 0} k \cot \delta(k) =-1/a_0$. While the result of the formal derivative expansion differs from the exact scattering length by about factor four, we find that the HAL QCD method gives the equivalent value with the exact one. This indicates that the choice of $k_{0}=0$ in the HAL QCD method gives a good description not only for the value of the phase shift (which is by definition $\delta=0$) but also for its slope at $k=0$.
\begin{table}
\caption{The scattering lengths from the local potentials by the formal derivative expansion (formal) and by the HAL QCD method with $k_0=0$ (HAL QCD), in comparison with the exact scattering length from the original nonlocal potential.
}
\label{tab_a0_compare}
\begin{ruledtabular}
\begin{tabular}{c c c c}
& formal & HAL QCD & exact \\ \hline
scattering length [fm] & 6.55 & 24.48 & 24.48 
\end{tabular}
\end{ruledtabular}
\end{table}

\begin{figure}[tbp]
    \centering
    \includegraphics[bb=0 0 1706 1250,width=0.45\textwidth]{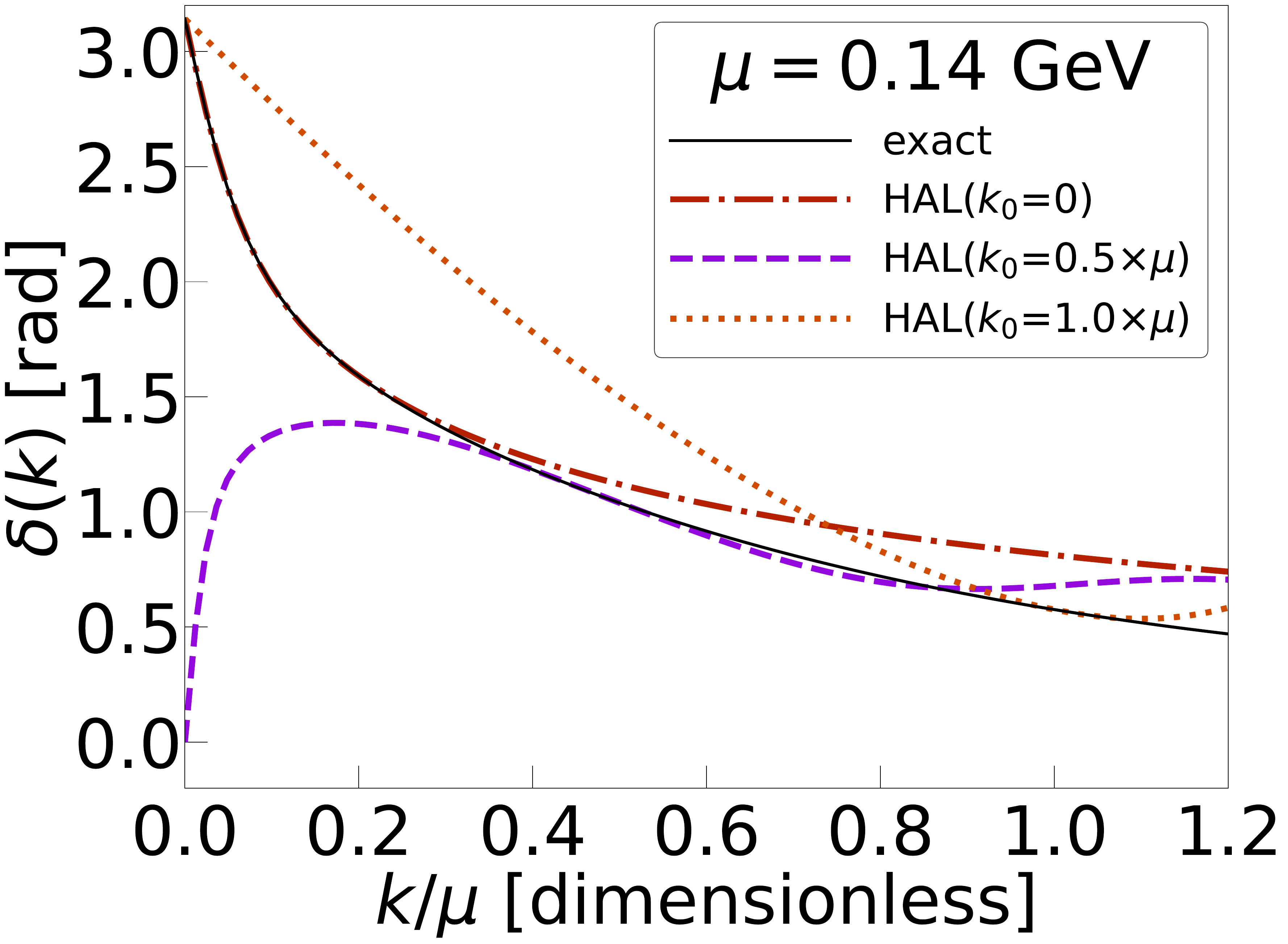}
    \caption{The phase shifts $\delta(k)$ from the potentials by the HAL QCD method with $k_0 = 0$ (dashed-dotted line), $k_0=\mu/2$ (dashed line), and $k_0=\mu$ (dotted line), as functions of the dimensionless momentum $k/\mu$ with the cutoff $\mu=0.14$ GeV in comparison with the exact phase shift (solid line).}
    \label{fig_comparedelta_Epot}
\end{figure}

It is worth examining the $k_0$ dependence of the local potential by the HAL QCD method again, from the perspective of the phase shift.
We show the phase shifts by the potentials with momentum $k_0=0, \mu/2$, and $\mu$, together with the exact one in Fig.~\ref*{fig_comparedelta_Epot}.
Although the potentials in Fig.~\ref*{fig_VHAL} do not show strong dependence on $k_0$, Fig.~\ref*{fig_comparedelta_Epot} demonstrates that the behavior of the phase shift drastically changes for different values of $k_{0}$.
This is because, the binding energy of $X(3872)$ is so small that even tiny change of the potential strength can make $X(3872)$ from bound to unbound.
In fact, we see that the system has a bound state for $k_0^2/2m =0$ and $E_{\mu}$ as indicated by the negative slope at $k=0$, but the bound state disappears for $k_0^2/2m=E_{\mu}/2$.
Table~\ref*{tab_a0_HAL} shows the detailed $k_0$ dependence of the scattering length $a_0$ of the potential by the HAL QCD method. We find from the table that the scattering length does not behave monotonically with respect to $k_{0}$. This reflects the fact that the HAL QCD method is formulated to reproduce the exact phase shift at $k=k_0$, but the phase shift is also fixed to be $\delta=0$ at $k=0$, and therefore the interpolation between $k=0$ and $k=k_{0}$ gives the strong $k_{0}$ dependence of the scattering length.
\begin{table}
    \caption{The $k_0$ dependence of the scattering length $a_0$ from the potential by the HAL QCD method, with $\mu=m_{\pi}=0.14$ GeV and $\mu=m_{\rho}=0.77$ GeV.}
    \label{tab_a0_HAL}
    \begin{ruledtabular}
    \begin{tabular}{c c c}
     $k_0/\mu$ [dimensionless] & $a_0(\mu=m_{\pi})$ [fm] & $a_0(\mu=m_{\rho})$ [fm] \\ \hline
    0   & 24.48 & 22.36\\
    0.1 & 24.14 & 8.32\\
    0.2 & 21.38 & 2.84\\
    0.3 & 22.68 & 1.34\\
    0.4 & 17.17 & 0.79\\
    0.5 & $-63.97$ & 0.71\\
    0.6 & 9.33 & 0.01\\
    0.7 & 5.88 & 0.23\\
    0.8 & $-0.78$ & 0.60\\
    0.9 & $-1.27$ & $-0.13$\\
    1   & 5.21 & $-0.20$\\
    \end{tabular}
    \end{ruledtabular}
    \end{table}

\begin{figure}[tbp]
    \centering
    \includegraphics[bb=0 0 1669 1249,width=0.45\textwidth]{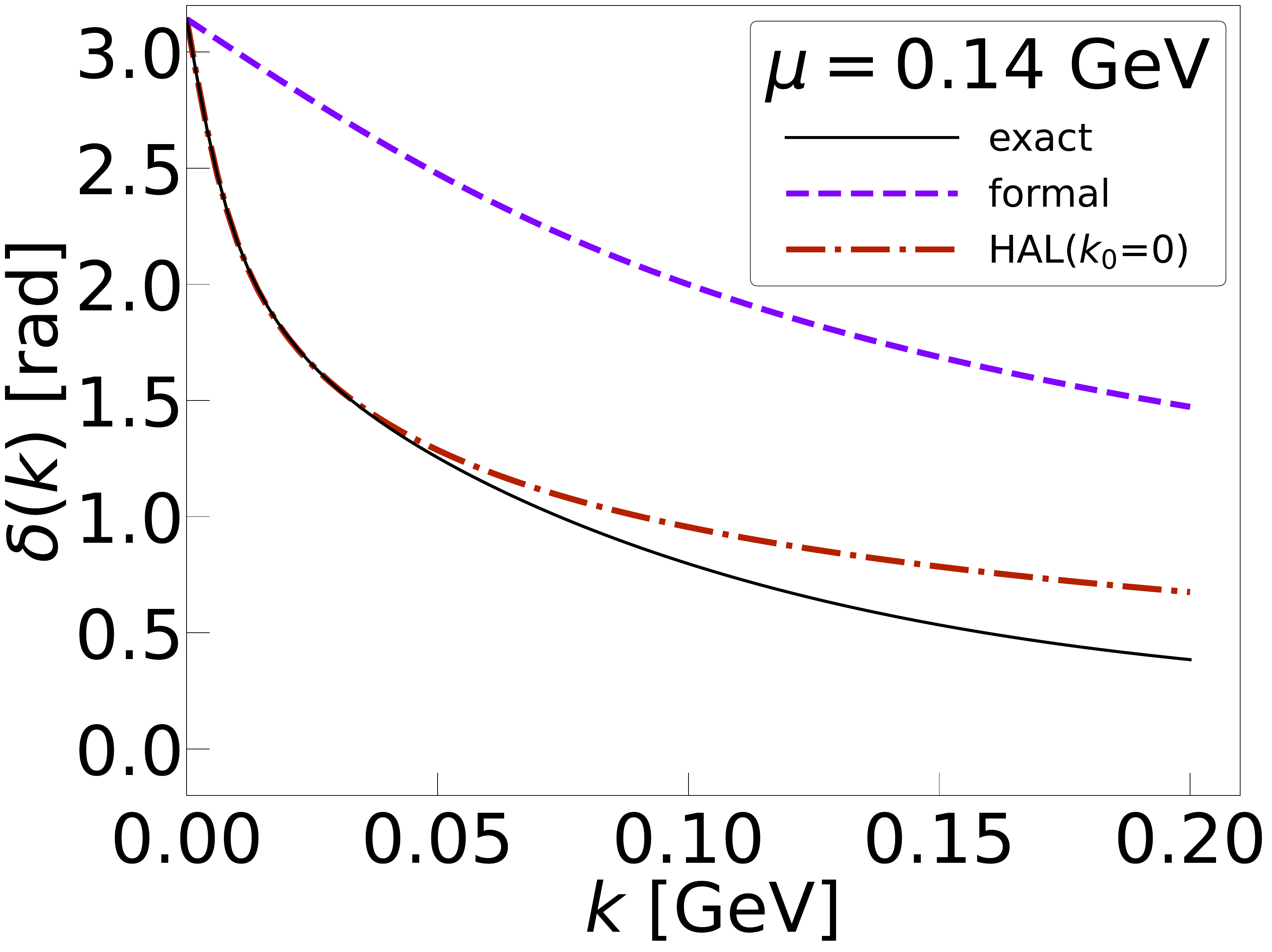}
    \caption{
    The phase shifts from the local potentials by the formal derivative expansion (dashed line) and by the HAL QCD method (dashed-dotted line) as functions of the momentum $k$ with the cutoff $\mu=0.14$ GeV in comparison with the exact phase shift (solid line).
    }
    \label{fig_compare_delta_k_mu014}
\end{figure}

\begin{figure}[tbp]
    \centering
    \includegraphics[bb=0 0 1669 1249,width=0.45\textwidth]{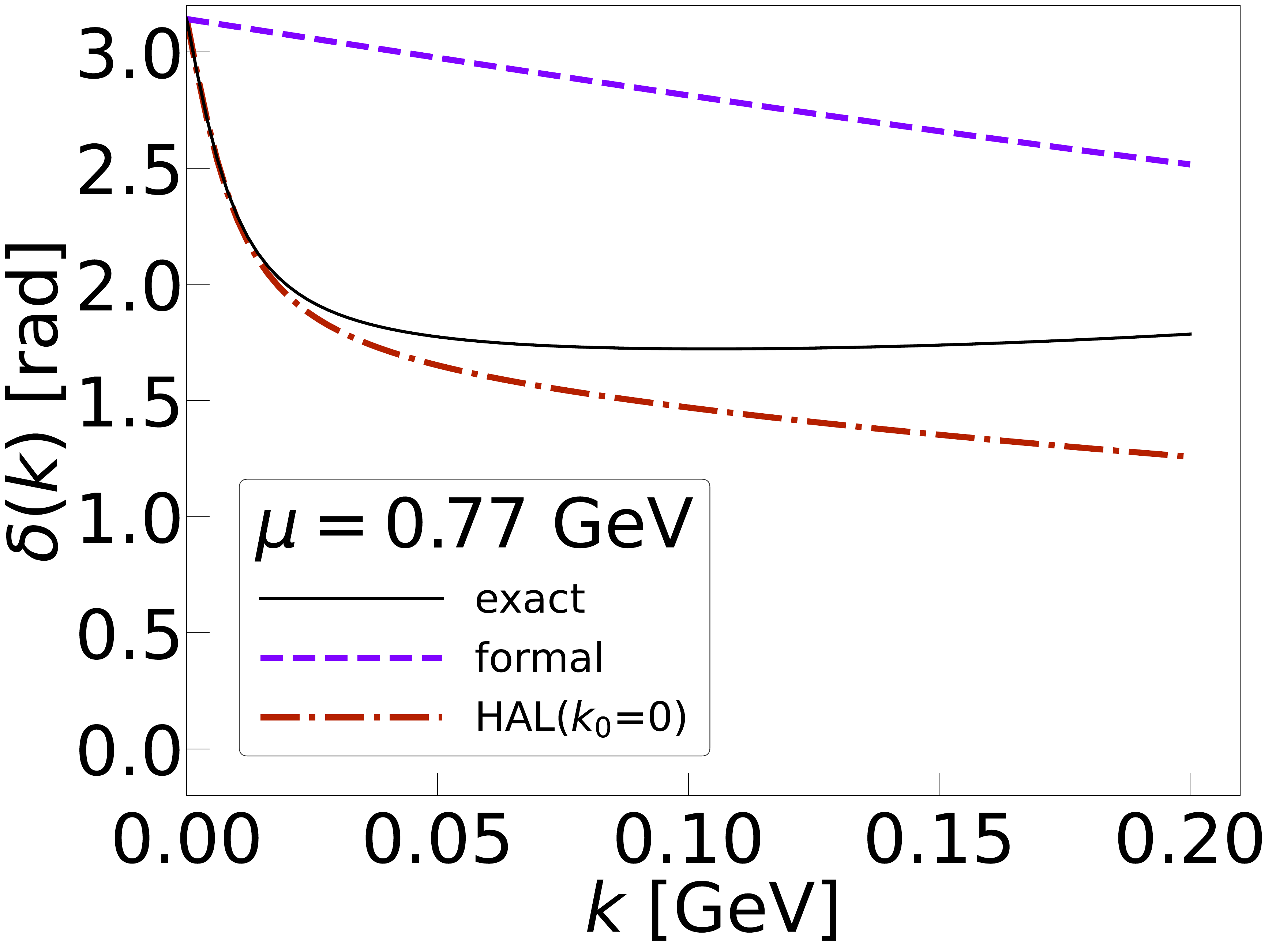}
    \caption{
    The same with Fig.~\ref{fig_compare_delta_k_mu014} but with the cutoff $\mu=0.77$ GeV. }
    \label{fig_compare_delta_k_mu077}
\end{figure}

Finally, we study the $\mu$ dependence of the approximation methods of the formal derivative expansion and the HAL QCD method.
Here, we vary only the cutoff $\mu$, with fixing the binding energy of $X(3872)$ as $B_{X(3872)} = 40$ keV and the momentum $k_0$ for the HAL QCD method as $k_0 = 0$.
We note that in this case the exact phase shift from the nonlocal potential depends on $\mu$ as indicated in Eq.~\eqref{eq_kcotdelta}.
We show the phase shifts as functions of momentum $k$ in Fig.~\ref*{fig_compare_delta_k_mu014} with $\mu=m_{\pi}=0.14$ GeV, and in Fig.~\ref*{fig_compare_delta_k_mu077} with $\mu=m_{\rho}=0.77$ GeV.
We see that the $k$ dependence of the phase shifts changes for $\mu=0.77$ GeV, including the exact one. 
We find from the figures that the HAL QCD method better reproduces the exact phase shift than the formal derivative expansion at low momentum region, irrespective of the value of the cutoff $\mu$.
We also show the $k_0$ dependence of the scattering length $a_0$ at $\mu=0.77$ GeV in Table~\ref*{tab_a0_HAL}.
As in the case of $\mu=0.14$ GeV, we see that the $k_0$ dependence of $a_0$ shows the nonmonotonic behavior also at $\mu=0.77$.
In this way, the properties of the local approximation discussed above does not depend on $\mu$. 

\section{Summary}\label{sc_summary}
In this paper, we have discussed the properties of the effective potentials with the channel coupling of the quark and hadron degrees of freedom. We examine the local approximation methods of the obtained effective potential using the model of $X(3872)$.

First, we have formulated the Hamiltonian that describes the coupled-channel problem between the quark and hadron degrees of freedom, and then have derived an effective potential that contains the channel coupling contribution in the Feshbach's method. As a result, it is found that the effective potential is obtained as an energy-dependent and nonlocal potential.
We show that the quark-antiquark potential develops an imaginary part due to the coupling to the two-hadron scattering states at finite quark masses.

In order to extract the physical mechanism of the interaction, we discuss the approximation of the nonlocal effective potential into a local one. In this study, we introduce two approximation methods: the formal derivative expansion and the HAL QCD method. 
Analytic expressions of the approximated local potentials by these methods are derived for an energy-dependent separable potential with the Yukawa form factor, which is a prototype of the effective nonlocal potentials.

Finally, we apply the framework of the hadron-hadron effective potential to $X(3872)$, one of the prominent examples of the exotic hadrons. We construct the nonlocal ${D^0}\bar{D}^{*0}$ potential coupled with the $\chi_{c1}(2P)$ bound state of $c\bar{c}$ to reproduce the mass of $X(3872)$. The approximated local potentials and the scattering phase shifts from the potentials are computed numerically. We show the quantitative deviation of the obtained local potentials and phase shifts for different approximation methods. Through the comparison with the exact solution obtained from the original nonlocal potential, we find that the HAL QCD method better reproduces the exact phase shift than the formal derivative expansion, in the presence of the energy dependence. However, it should be noted that the observable quantities such as the phase shift and scattering length are sensitive to the changes of the momentum $k_0$ for the system with a shallow bound state like $X(3872)$, although the approximated local potential has a weak $k_{0}$ dependence.

As future prospects, it is important to apply the present results to more realistic models of $X(3872)$. For example, the coupling to the charged $D^{\pm}D^{*\mp}$ channel and to the decay channels of $J/\psi\rho$ and $J/\psi\pi\pi$, the inclusion of the additional $c\bar{c}$ bound states, and the use of the $^3P_{0}$ model for the transition potential $V^t$ will improve the theoretical model of $X(3872)$. While we have focused on the hadron-hadron effective potentials for the application, the framework presented here can also be used to investigate the influence of the hadron degrees of freedom on the quark-antiquark effective potentials. In addition, the application to the other exotic hadrons, such as the $H$-dibaryon in the $\Lambda\Lambda$-$N\Xi$ system~\cite{Jaffe:1976yi,Yamaguchi:2016kxa,HALQCD:2019wsz} and $T_{cc}$ in the $DD^{*}$ scattering~\cite{LHCb:2021vvq,LHCb:2021auc}, can be pursued in the framework of the present work.

\begin{acknowledgments}
    The authors thank Sinya Aoki, Atsushi Hosaka, Akira Ohnishi and Shoichi Sasaki for useful discussions on the coupling of the quark and hadron potentials during the YITP workshop YITP-T-14-03 on “Hadrons and Hadron Interactions in QCD” held at Yukawa Institute for Theoretical Physics, Kyoto University.
    This work has been supported in part by the Grants-in-Aid for Scientific Research from JSPS (Grants
    No. JP22K03637, 
    No. JP19H05150, 
    No. JP18H05402). 
    This work was supported by JST, the establishment of university fellowships towards the creation of science technology innovation, Grant No. JPMJFS2139. 
\end{acknowledgments}

%

\end{document}